\documentclass[preprint2]{aastex}
\usepackage{spr-astr-addons}
\usepackage{graphicx}
\begin{document}
\title{GRMHD/RMHD Simulations \& Stability of Magnetized Spine-Sheath
  Relativistic Jets}
\shorttitle{Spine-Sheath Jet Stability}       
\shortauthors{Hardee et al.}
\author{Philip Hardee\altaffilmark{1}, Yosuke
Mizuno\altaffilmark{2,3}, \& Ken-Ichi Nishikawa\altaffilmark{2,4}}
%


\altaffiltext{1}{The University of Alabama (UA), Tuscaloosa, AL, 35487
USA}

\altaffiltext{2}{National Space Science \& Technology Center (NSSTC),
Huntsville, AL 35805 USA}

\altaffiltext{3}{NASA/Marshall Space Flight Center (NASA/MSFC),
Huntsville, AL 35805 USA}

\altaffiltext{4}{University of Alabama at Huntsville (UAH),
Huntsville, AL 35805 USA}

\begin{abstract}
A new general relativistic magnetohydrodynamics ({\bf GRMHD}) code
``RAISHIN'' used to simulate jet generation by rotating and
non-rotating black holes with a geometrically thin Keplarian accretion
disk finds that the jet develops a spine-sheath structure in the
rotating black hole case.  Spine-sheath structure and strong magnetic
fields significantly modify the Kelvin-Helmholtz ({\bf KH}) velocity
shear driven instability. The RAISHIN code has been used in its
relativistic magnetohydrodynamic ({\bf RMHD}) configuration to study
the effects of strong magnetic fields and weakly relativistic sheath
motion, $c/2$, on the KH instability associated with a relativistic,
$\gamma = 2.5$, jet spine-sheath interaction. In the simulations sound
speeds up to $ \sim c/\sqrt 3$ and Alfv\'en wave speeds up to $\sim
0.56~c$ are considered.  Numerical simulation results are compared to
theoretical predictions from a new normal mode analysis of the RMHD
equations.  Increased stability of a weakly magnetized system
resulting from $c/2$ sheath speeds and stabilization of a strongly
magnetized system resulting from $c/2$ sheath speeds is found.
\end{abstract}


\vspace{-0.5cm}
\section{Introduction}
\label{sec:intro}
\vspace{-0.2cm}

Relativistic jets are associated with active galactic nuclei and
quasars ({\bf AGN}), with black hole binary systems ({\bf
microquasars}), and are thought responsible for the gamma-ray bursts
({\bf GRBs}). The observed proper motions in AGN and microquasar jets
imply speeds from $\sim 0.9~c$ (e.g., Mirabel \& Rodriquez 1999) up to
$\sim 0.999~c$ (e.g., the 3C\,345 jet Zensus et al.\ 1995; Steffen et
al.\ 1995), and the inferred speeds for GRBs are $\sim 0.99999~c$
(e.g., Piran 2005).

Jets at the larger scales may be kinetically dominated and contain
relatively weak magnetic fields, but stronger magnetic fields exist
closer to the acceleration and collimation region. Here GRMHD
simulations of jet formation (e.g., Koide et al.\ 2000; Nishikawa et
al.\ 2005; De Villiers et al.\ 2003, 2005; Hawley \& Krolik 2006;
McKinney \& Gammie 2004; McKinney 2006; Mizuno et al.\ 2006) and
earlier theoretical work (e.g., Lovelace 1976; Blandford 1976;
Blandford \& Znajek 1977; Blandford \& Payne 1982) invoke strong
magnetic fields. Additionally, the GRMHD simulations suggest that jets
driven by magnetic fields threading the ergosphere can reside within a
broader sheath outflow driven by the magnetic fields anchored in the
accretion disk (e.g., McKinney 2006; Hawley \& Krolik 2006; Mizuno et
al.\ 2006), or less collimated accretion disk wind (e.g., Nishikawa et
al.\ 2005).

Recent observations of QSO winds with speeds, $\sim 0.1 - 0.4c$, also
indicate that a jet could reside in a high speed sheath (Chartas et
al.\ 2002, 2003; Pounds et al.\ 2003a, 2003b; Reeves et al.\
2003). Circumstantial evidence such as the requirement for large
Lorentz factors suggested by the TeV BL Lacs when contrasted with much
slower observed motions has been used to suggest the presence of a
spine-sheath morphology (Ghisellini et al.\ 2005), and Siemignowska et
al.\ (2007) have proposed a spine-sheath model for the PKS 1127-145
jet.  Spine-sheath structure has also been proposed based on
theoretical arguments (e.g., Sol et al.\ 1989; Henri \& Pelletier
1991; Laing 1996; Meier 2003) and has been investigated in the context
of GRB jets (e.g., Rossi et al.\ 2002; Lazzatti \& Begelman 2005;
Zhang et al.\ 2003, 2004; Morsony et al.\ 2006).

In \S 2 we illustrate the spine-sheath configuration found by our
GRMHD jet generation simulations. Previous relativistic fluid
dynamical (RHD) simulation and theoretical work has shown
the importance of spine-sheath structure to KH instability (Hardee \&
Hughes 2003). In \S 3 we report on numerical results that extend this
previous investigation numerically and in \S 4 theoretically to the
strongly magnetized RMHD regime.

\vspace{-0.6cm}
\section{GRMHD Jet Spine-Sheath Generation}
\label{sec:1}
\vspace{-0.3cm}

In order to study the formation of relativistic jets from a
geometrically thin Keplerian disk, we use a 2.5-dimensional GRMHD
code with Boyer-Lindquist coordinates $(r, \theta, \phi)$. The
method is based on a 3+1 formalism of the general relativistic
conservation laws of particle number and energy momentum, Maxwell
equations, and Ohm's law with no electrical resistance (ideal MHD
condition) in a curved spacetime. In the
simulations presented here we use minmod slope limiter
reconstruction, HLL approximate Riemann solver, flux-CT scheme and
Noble's 2D method (see Mizuno et al.\ 2006 and references therein).

A geometrically thin Keplerian disk rotates around a black hole
(non-rotating, $a=0.0$ or rapidly co-rotating, $a=0.95$, here $a$ is
black hole spin parameter), where the disk density is 100 times the
coronal density. The thickness of the disk is $H/r \sim 0.06$. The
background corona is free-falling, and the initial magnetic field is
uniform and parallel to the rotational axis.  Simulations are
normalized by the speed of light, $c$, and the Schwarzschild radius,
$r_{\rm S}$, with timescale, $\tau_{\rm S} \equiv r_{\rm S}/c$. Values
of the magnetic field strength and gas pressure depend on the
normalized density, $\rho_{0}$. In these simulations the magnetic
field strength, $B_{0}$, is set to $0.05 \sqrt{\rho_{0} c^{2}})$. The
$128 \times 128$ computational grid with logarithmic spacing in the
radial direction spans the region $1.1 r_{\rm S} \le r \le 20.0 r_{\rm
S}$ (non-rotating black hole) and $0.75 r_{\rm S}\le r \le 20.0 r_{\rm
S}$ (rapidly rotating black hole) and $0.03 \le \theta \le \pi/2$
where we assume axisymmetry with respect to the $z$-axis and mirror
symmetry with respect to the equatorial plane. We employ a free
boundary condition at the inner and outer boundaries in the radial
direction.

Figure 1 shows snapshots of the density (panels (a) and (b)), plasma
beta ($\beta=p_{\rm gas}/p_{\rm mag}$) distribution (panels (c) and
(d)), and total velocity (panels (e) and (f)) for the non-rotating
black hole, $a=0.0$ (left panels); and the rapidly rotating black
hole, $a=0.95$ (right panels); at each simulation's terminal time
(non-rotating: $t = 275\tau_{\rm S}$ and rotating: $t = 200\tau_{\rm
S}$). At the marginally stable circular orbit ($r = 3r_{\rm S}$) the
disk orbits the black hole in about $40 \tau_{\rm S}$. The total
velocity distribution of non-rotating and rapidly rotating black hole
cases are shown in Figs. 1e and 1f.
\begin{figure}[h!]
\vspace{-0.1cm}
\begin{center}
\includegraphics[width=0.48\textwidth]{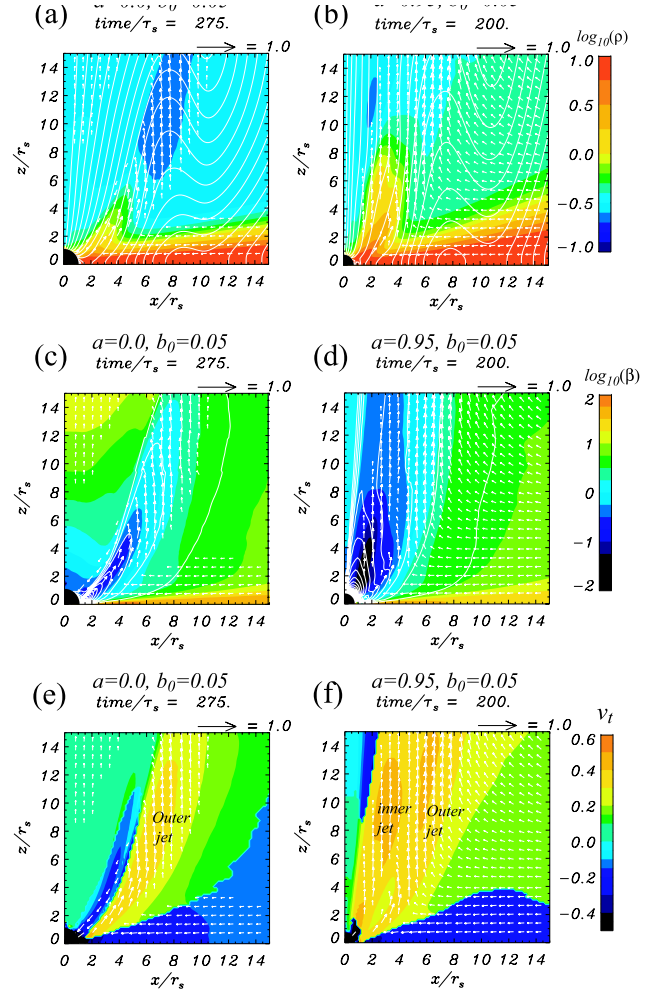}
\end{center}
\vspace{-0.55cm}
\caption{Snapshots of the non-rotating black hole ({\it a, c, e}) and
the rapidly rotating black hole ({\it b, d, f}) at the applicable
terminal simulation time. The color scales show the logarithm of
density (upper panels), plasma beta ($\beta = p_{\rm gas} / p_{\rm
mag}$; middle panels) and total velocity (lower panels). A negative
velocity indicates inflow towards the black hole. The white lines
indicate magnetic field lines (contour of the poloidal vector
potential; upper panels) and contours of the toroidal magnetic field
strength (middle panels). Arrows depict the poloidal velocities
normalized to light speed, as indicated above each panel by the
arrow.\vspace{-0.3cm}}
\label{fig:1}
\end{figure}
The jets in both cases have speeds greater than $0.4~c$ (mildly
relativistic) that are comparable to the Alfv\'{e}n speeds. In the
jets, toroidal velocity is the dominant velocity component. In the
rapidly rotating black hole case, the velocity distribution indicates
a two-component jet with the inner jet not seen in the non-rotating
black hole case. The inner jet is faster than the outer jet (over
$0.5c$).

\vspace{-0.6cm}
\section{RMHD Spine-Sheath Simulations}
\label{sec:2}
\vspace{-0.3cm}

In these simulations a ``preexisting'' jet is established across a
computational domain of $6 R_{j} \times 6 R_{j} \times 60 R_{j}$ with
$60 \times 60 \times 600$ zones. The jet is in total pressure balance
with a lower-density magnetized sheath with $\rho_{j}/\rho_{e}=2.0$,
where $\rho$ is the mass density in the proper frame. The jet speed is
$u_{j}=0.9165~c$ and $\gamma_j \equiv (1 - u_j^{2})^{-1/2}=2.5$. The
initial magnetic field is uniform and parallel to the jet flow.  A
precessional perturbation is applied at the inflow by imposing a
transverse component of velocity with $u_{\bot}=0.01u_{j}$.
Here we show simulations with a precessional perturbation of angular
frequency $\omega R_{j}/u_{j}=0.93$.  In order to investigate the
effect of an external wind, we have performed a no wind case
($u_{e}=0$) and a relativistic wind case ($u_{e}=0.5~c$).  Simulations
are halted after $\sim 60$ light crossing times of the jet radius (see
Mizuno et al.\ 2007 for details).

We have performed weakly magnetized simulations with sound speeds
$a_{e} \sim 0.57~c$ and $a_{j} \sim 0.51~c$, and Alfv\'{e}n speeds
$v_{Ae} \sim 0.07~c$ and $v_{Aj} \sim 0.06~c$.  The stabilizing effect
of a sheath wind is revealed in Figure 2.  Here we see considerable
reduction
\begin{figure}[h!]
\vspace{-0.1cm}
\begin{center}
\includegraphics[width=0.48\textwidth]{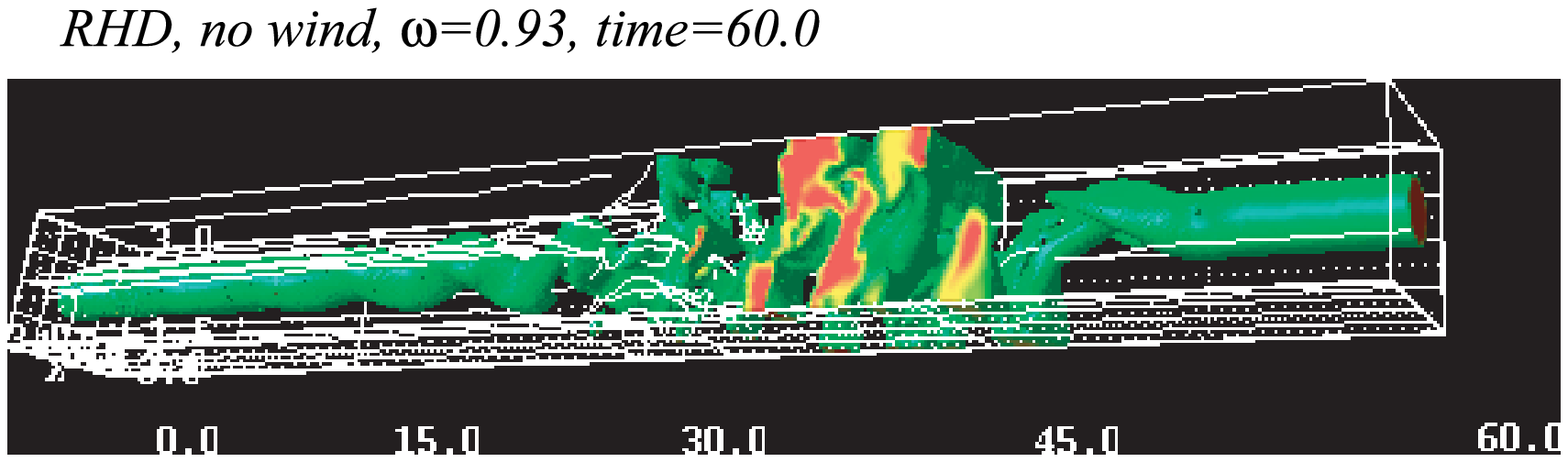}
\vspace{-0.3cm}

\includegraphics[width=0.48\textwidth]{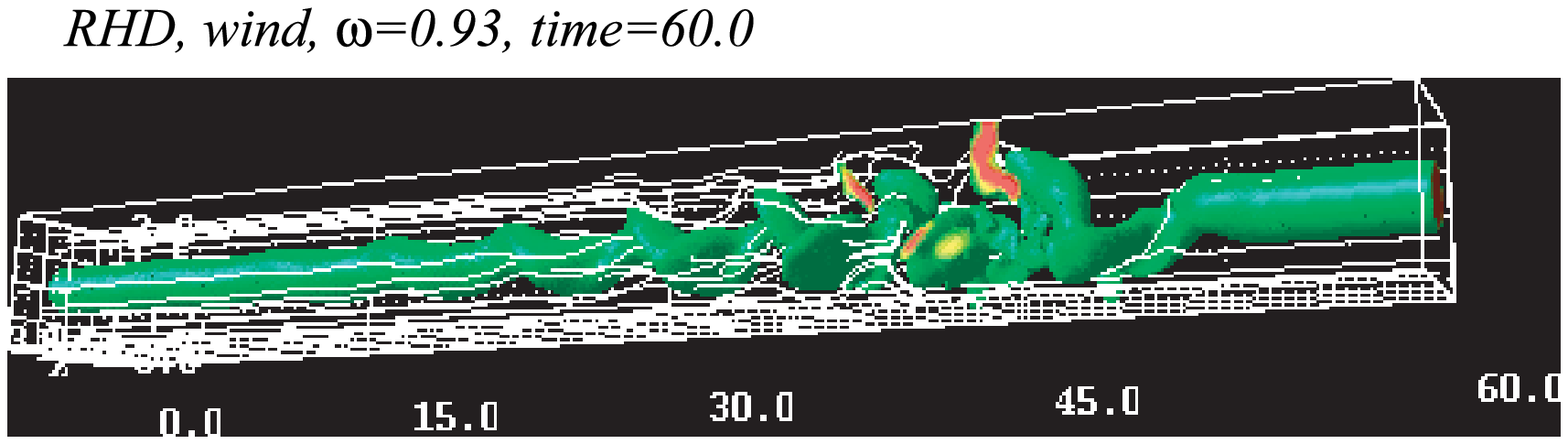} 
\end{center}
\vspace{-0.4cm}
\caption{3D isovolume density image of the weakly magnetized case with
no wind (top) and a c/2 wind (bottom). Magnetic field lines in
white. \vspace{-0.3cm}}
\label{fig:2a}
\end{figure}
in transverse structure and the jet spine reaches a larger distance
before disruption in the presence of a wind.

We have also performed strongly magnetized simulations with Alfv\'{e}n
speeds $v_{Ae} \sim 0.56~c$ and $v_{Aj} \sim 0.45~c$, and sound speeds
$a_{e} \sim 0.30~c$ and $a_{j} \sim 0.23~c$. The stabilizing influence
of a magnetic field and the stabilization of the jet spine in the
presence of a magnetized sheath wind is shown in Figure 3.
\begin{figure}[h!]
\vspace{0.1cm}
\begin{center}
\includegraphics[width=0.48\textwidth]{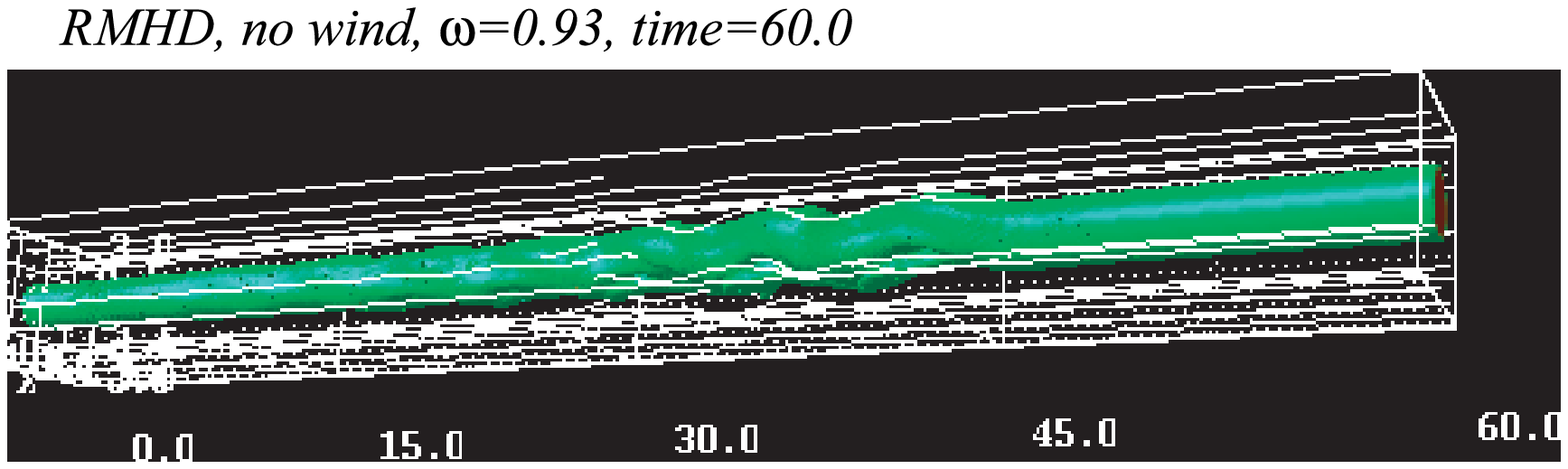}
\vspace{-0.3cm} 

\includegraphics[width=0.48\textwidth]{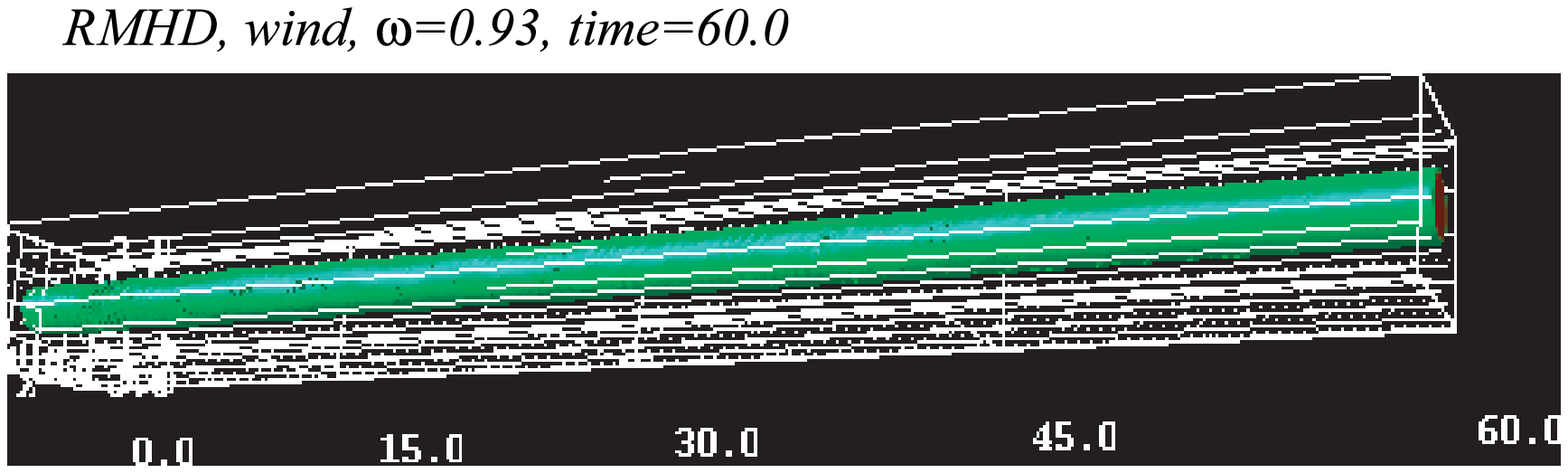} 
\end{center}
\vspace{-0.4cm} 
\caption{3D isovolume density image of the strongly magnetized case
with no wind (top) and a c/2 wind (bottom). Magnetic field lines in
white. \vspace{-0.3cm}}
\label{fig:3}
\end{figure}
Here we see that the presence of the strong magnetic field
has stabilized the jet spine even more than occured for the weakly
magnetized wind case and the initial helical perturbation is damped in
the presence of the strongly magnetized sheath wind.

More quantitatively we can analyse the growth or damping of the
initial perturbation via 1D cuts in the radial velocity as shown in
Figure 4.
\begin{figure}[h!]
\vspace{0.2cm}
\begin{center}
\includegraphics[width=0.48\textwidth]{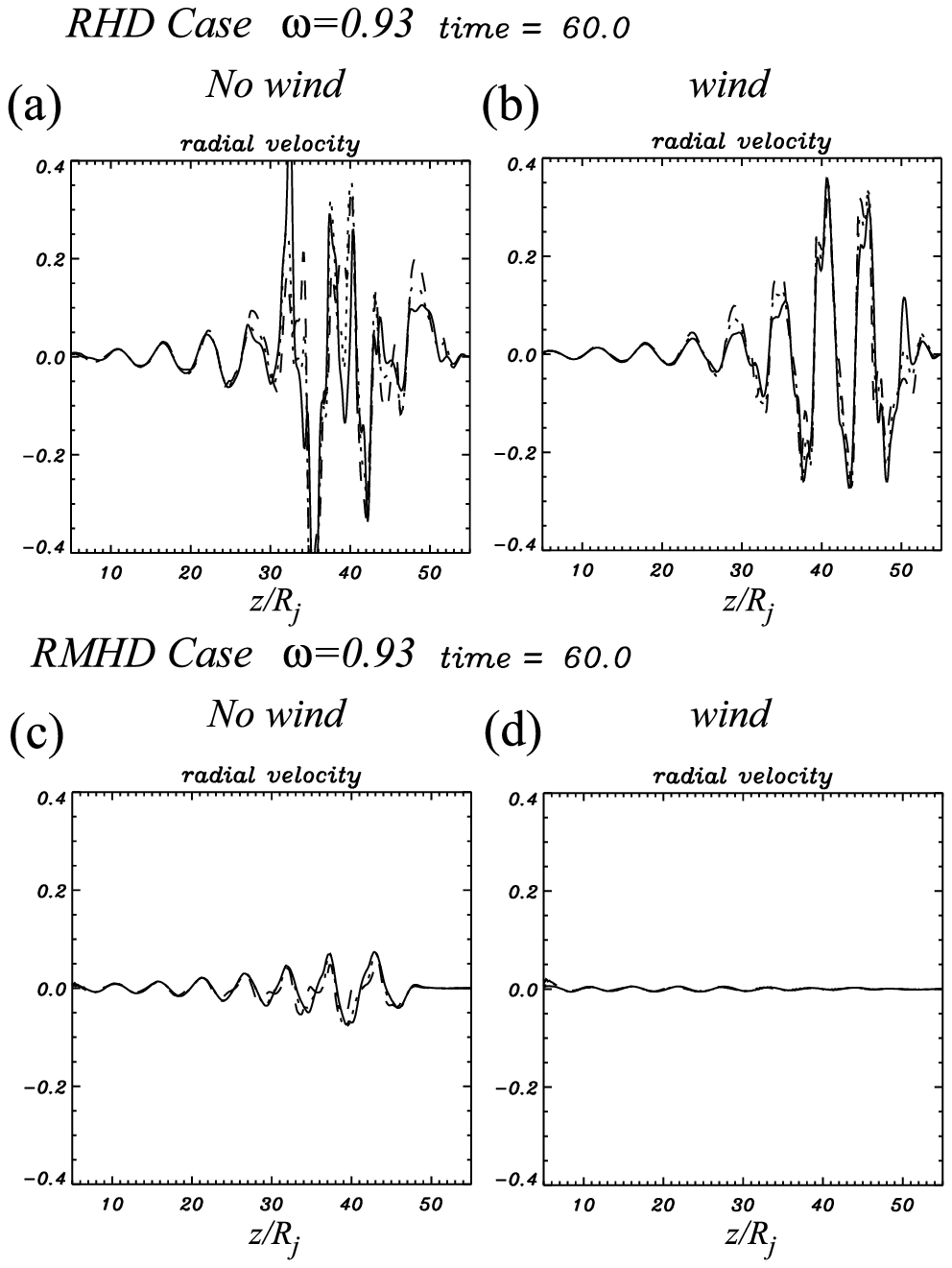}
\end{center}\vspace{-0.5cm}
\caption{Radial velocity ($v_x$) along one dimensional cuts parallel
to the jet axis and located at $x/R_J =$ 0.2 (solid line), 0.5 (dotted
line) and 0.8 (dashed line) for the weakly magnetized (top) and
strongly magnetized cases (bottom).\vspace{-0.3cm}}
\label{fig:4}
\end{figure}
Measurable reduction in transverse motion is seen for the weakly
magnetized wind case, significant reduction occurs for
the strongly magnetized no wind case and stabilization occurs in
the strongly magnetized wind case.

\vspace{-0.5cm}
\section{RMHD Spine-Sheath Stability}
\label{sec:3}
\vspace{-0.3cm}

Stability of a jet spine-sheath configuration can be analyzed by
modeling the jet/spine as a cylinder of radius R embedded in an
infinite sheath. A dispersion relation describing the growth or
damping of the normal modes can be derived assuming uniform conditions
within the spine, e.g., a uniform proper density, $\rho _{j}$, axial
magnetic field, $B_{j}=B_{j,z}$, and velocity,
$\mathbf{u}_{j}=u_{j,z}$, and assuming uniform conditions in the
external sheath, e.g., a uniform proper density, $ \rho _{e}$, axial
magnetic field, $B_{e}=B_{e,z}$, and velocity $\mathbf{u}_{e}=u_{e,z}$
(see Hardee 2007).

Each normal mode consists of a single fundamental and multiple body
wave solutions to the dispersion relation. In the low frequency limit
the helical fundamental mode has an analytical solution given by
\vspace{-0.1cm}
\begin{equation}
\frac{\omega }{k}=\frac{\left[ \eta u_{j}+u_{e}\right] \pm i\eta
^{1/2}\left[ \left( u_{j}-u_{e}\right) ^{2}-V_{As}^{2}/\gamma
_{j}^{2}\gamma _{e}^{2} \right] ^{1/2}}{(1+V_{Ae}^{2}/\gamma
_{e}^{2}c^{2})+\eta (1+V_{Aj}^{2}/\gamma _{j}^{2}c^{2})} 
\label{eq1}
\vspace{-0.3cm}
\end{equation}
where $\eta \equiv \gamma _{j}^{2}W_{j}\left/ \gamma
_{e}^{2}W_{e}\right.$, $V_{A}^2 \equiv B^2/4 \pi W$, $W\equiv \rho
+\left[ \Gamma /\left( \Gamma -1\right) \right] P/c^{2}$ is the
enthalpy, and
\vspace{-0.1cm}
\begin{equation}
V_{As}^{2}\equiv \left( \gamma _{Aj}^{2}W_{j}+\gamma
_{Ae}^{2}W_{e}\right) \frac{B_{j}^{2}+B_{e}^{2}}{4\pi W_{j}W_{e}}~.
\label{eq2}
\vspace{-0.3cm}
\end{equation}
In equation (2) $\gamma _{Aj,e}\equiv (1-v_{Aj,e}^{2}/c^{2})^{-1/2}$
is the Alfv\'{e}n Lorentz factor.  The jet is stable when
\vspace{-0.1cm}
\begin{equation}
\left( u_{j}-u_{e}\right)^{2}-V_{As}^{2}/\gamma _{j}^{2}\gamma
_{e}^{2} < 0~.
\label{eq3}
\vspace{-0.3cm}
\end{equation}
In the low frequency limit the real part of the first helical
body mode has an analytical solution given by
\vspace{-0.2cm}
\begin{equation}
kR\approx k^{\min }R\equiv \frac{5}{4} \pi \left[ \frac{
v_{msj}^2u_{j}^{2}-v_{Aj}^2a_{j}^{2}}{\gamma
_{j}^{2}(u_{j}^{2}-a_{j}^{2})(u_{j}^{2}-v_{Aj}^{2})}\right]^{1/2}.
\label{eq4}
\vspace{-0.3cm}
\end{equation}
In equation (4) $v_{ms}$ is a magnetosonic speed defined by $v_{ms}
\equiv \left[a^2/\gamma_A^2 + v_A^2\right]^{1/2}$ where $a$ is the
sound speed, and $v_{A}$ is the Alfv\'{e}n wave speed.

Equations (1 \& 4) provide estimates for the helical fundamental and
first body modes that can be followed by root finding techniques to
higher frequencies.  The results of numerical solution to the
dispersion relation for the parameters appropriate to the numerical
simulations shown in \S 3 are displayed in Figure 5.

\begin{figure}[h!]
\vspace{0.1cm}
\begin{center}
\includegraphics[width=0.48\textwidth]{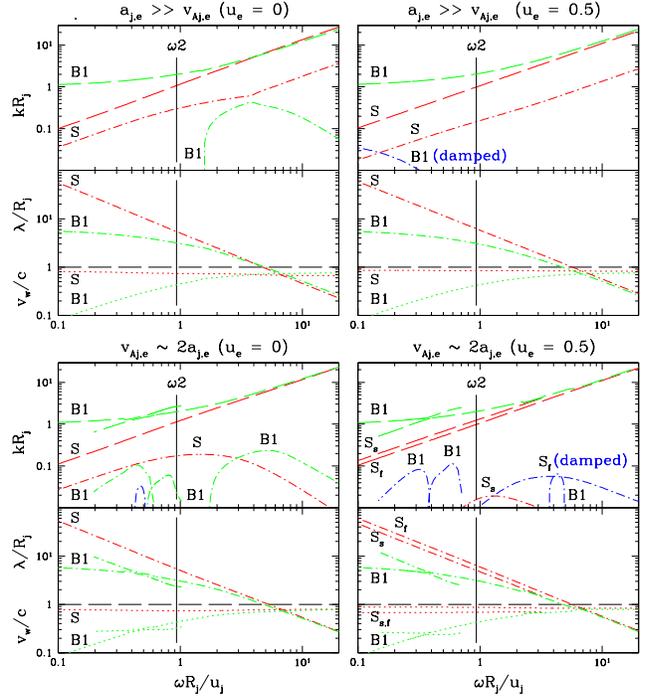}
\end{center}
\vspace{-0.5cm}
\caption{Solutions for helical fundamental (red lines) and first body
(green lines) modes for weakly magnetized ($a_{j,e} \gg v_{Aj,e}$) and
strongly magnetized ($ v_{Aj,e} \sim 2a_{j,e}$) jet simulations with
no wind ($u_e = 0$) and with a c/2 wind ($u_e = 0.5$).  Solutions show
the real, $k_rR_j$, (dashed lines) and imaginary, $k_iR_j$, (dash-dot
lines) parts of the wavenumber as a function of the angular frequency,
$\omega R_j/ u_j$. Where the imaginary part of the wavenumber is shown
in blue, the solution is damped.  Immediately under the solutions for
fundamental ({\bf S}) and first body ({\bf B1}) modes is a panel that
shows the wavelength, $\lambda/R_j$, (dash-dot lines) and wave speed,
$v_w/c$, (dotted lines).  The simulation precession frequency $\omega
2 = 0.93$ is indicated by the vertical solid line. \vspace{-0.3cm}}
\label{fig:5}
\end{figure}

In the weakly magnetized cases fundamental ({\bf S}) mode solutions
consist of a growing (shown) and damped (not shown) solution pair (see
eq.\ 1) and first body ({\bf B1}) mode solutions consist of a real and
growing or damped solution pair. The presence of the external wind
flow leads to reduced growth of the S mode and weak damping of the B1
mode.

In the strongly magnetized no wind case S mode solutions again consist
of a growing and damped solution pair.  However, we now find multiple
growing solutions associated with the B1 mode at lower frequencies,
and a modest damping rate accompanies the crossing of the multiple
body mode solutions.  At higher frequencies the B1 mode is similar to
the weakly magnetized case.

In the strongly magnetized wind case weak growth is associated with
the slower, $S_{s}$, moving shorter wavelength solution and weak
damping is associated with the faster, $S_{f}$, moving longer
wavelength solution.  At frequencies, $\lesssim \omega 2$, the growth
rate is larger than the damping rate but at higher frequencies the
damping rate is larger than growth rate for the S mode solution pair.
In general the B1 mode is damped.

\vspace{-0.6cm}
\section{Conclusions}
\label{sec:concld}
\vspace{-0.3cm}

Increased stability of the weakly-magnetized system with mildly
relativistic sheath flow and stabilization of the strongly-magnetized
system with mildly relativistic sheath flow is in agreement with
theoretical results.  In the fluid limit the present results confirm
earlier results obtained by Hardee \& Hughes (2003), who found that
the development of sheath flow around a relativistic jet spine
explained the partial stabilization of the jets in their numerical
simulations.

The simulation results agree with theoretically predicted wavelengths
and wave speeds. On the other hand, growth rates and spatial growth
lengths obtained from the linearized equations or from the present
relatively low resolution simulations only provide guidelines to the
rate at which perturbations grow or damp.

A rapid decline in perturbation amplitudes in the sheath as a function
of radius, governed by a Hankel function in the dispersion relation,
suggests that the present results will apply to sheaths more than about
three times the spine radius in thickness.

Where flow and magentic fields are parallel, current driven ({\bf CD})
modes are stable (Isotomin \& Pariev 1994, 1996). However, we expect
magnetic fields to have a significant toroidal component. Provided
radial gradients are not too large we expect the present results to
remain approximately valid where $u_{j,e}$ and $B_{j,e}$ refer to
poloidal velocity and magnetic field components.

In the helically twisted magnetic and flow field regime likely to be
relevant to many astrophysical jets CD modes (Lyubarskii 1999) and/or
KH modes could be unstable.  While both CD and KH instability produce
helically twisted structure, the conditions for instability, the radial
structure, the growth rate and the pattern motions are different. These
differences may serve to identify the source of helical structure on
relativistic jet flows and allow determination of jet properties near
to the central engine.

\vspace{-0.2cm}
\begin{acknowledgements}
Resarch supported by NASA/MSFC cooperative agreement NCC8-256 and NSF
award AST-0506666 to UA (P.\ Hardee), the NASA/MSFC postdoctoral
program administered by ORAU (Y.\ Mizuno), and by NASA awards
NNG-05GK73G, HST-AR-10966.01-A and NSF award AST-0506719 to UAH (K.\
Nishikawa). The numerical simulations were performed on the IBM p690
at NCSA and the Altix3700 BX2 at YITP.
\end{acknowledgements}


\end{document}